\documentclass[twocolumn,showpacs,showkeys]{revtex4}
\usepackage{graphicx}
\usepackage{bm}
\usepackage{color}
\usepackage{amsmath}
\usepackage{natbib}

\begin{document}

\title{Bosonic bright soliton in the mixture of repulsive Bose-Einstein condensate
and polarized ultracold fermions under influence of the pressure evolution}

\author{Pavel A. Andreev}
\email{andreevpa@physics.msu.ru}
\affiliation{Department of General Physics, Faculty of physics, Lomonosov Moscow State University, Moscow, Russian Federation, 119991.}

\author{K. V. Antipin}
\email{kv.antipin@physics.msu.ru}
\affiliation{Department of Quantum Statistics and Field Theory,
Faculty of physics, Lomonosov Moscow State University, Moscow, Russian Federation, 119991.}

\author{Mariya Iv. Trukhanova}
\email{mar-tiv@yandex.ru}
\affiliation{Department of General Physics, Faculty of physics, Lomonosov Moscow State University, Moscow, Russian Federation, 119991.}
\affiliation{Russian Academy of Sciences, Nuclear Safety Institute (IBRAE), B. Tulskaya 52, Moscow, Russian Federation, 115191.}

\date{\today}

\begin{abstract}
Repulsive Bose-Einstein condensate,
where the short-range interaction is included up to the third order by the interaction radius,
demonstrates existence of a bright soliton in a narrow interval of parameters.
This soliton is studied here for the boson-fermion mixture,
where spin-1/2 fermions are consider in the regime of full spin polarization.
Influence of fermions via the boson-fermion interaction
is considered up to the third order by the interaction radius.
Fermions themselves are considered by hydrodynamic model including the pressure evolution equation.
Interaction between fermions is considered.
The first order by the interaction radius gives zero contribution in the Euler equation and the pressure evolution equation,
but the third order by the interaction radius provides nonzero contributions in both equations.
Repulsive (attractive) boson-fermion interaction leads to the bright (dark) fermionic soliton.
\end{abstract}

%Use showkeys class option if keyword

%03.75.Hh,
%03.75.Kk,
%67.85.Pq

\maketitle

%%%%%%%%%%TEXT

\section{Introduction}

Solitons \cite{Yefsah Nat 13}, \cite{Becker Nat Phys 08}, \cite{Tylutki NJP 16},
vorticities \cite{Anderson PRL 01}, \cite{Guenther PRA 17}, \cite{Gautam PRA 14},
skyrmions \cite{Lee Sc Ad 18} are fundamental nonlinear excitations in quantum gases.
The quantum droplets formation is also detected
in Bose-Einstein condensate of rare-earth atoms
due to the dipole-dipole interaction and quantum correlations \cite{Wachtler 1601}, \cite{Bisset PRA 16}, \cite{Wachtler 1605}.
This work is focused on solitons in boson-fermion mixtures  \cite{Karpiuk PRL 04}
studied in terms of hydrodynamic model.
The boson-fermion mixtures are experimentally obtained in different combinations, such as
$^{7}$Li-$^{6}$Li \cite{Truscott Science 01}, \cite{Schreck PRL 02},
$^{23}$Na-$^{6}$Li \cite{Hadzibabic PRL 02},
and $^{87}$Rb-$^{40}$K \cite{Roati PRL 02}.
If Bose-Einstein condensate (BEC) is considered in terms of the Gross-Pitaevskii equation \cite{Dalfovo RMP 99}
it reveals the bright soliton at the attraction between bosons
or the dark soliton in the repulsive BECs.
Noninteracting degenerate fermions also demonstrate existence of soliton \cite{Karpiuk JPB 02},
where pair of a bright and dark solitons for trapped fermions is obtained.
However, the extended model with nonlocal short-range interaction considered up to the third order by the interaction radius
leads to formation of a bright soliton (earlier called the bright-like soliton) in repulsive BECs \cite{Andreev MPL B 12}.
The bright soliton in repulsive BECs is studied for the anisotropic short-range interaction \cite{Andreev LP 19},
which is also considered up to the third order by the interaction radius.

Fermionic bright solitons at attractive boson-fermion interaction are found in Refs.
\cite{Adhikari PRA 07}, \cite{Parker PRA 12}.
Fermionic dark solitons are obtained in \cite{Truscott Science 01}.
The dark solitons in the Fermi superfluid within the Bogoliubov-de Gennes theory of the BEC to BCS crossover
using the random-phase approximation are considered in \cite{Cetoli PRA 13},
where the decay of the soliton via the snake instability is studied.

BEC-BCS crossover in the two component fermion systems is actively studying branch of the field of ultracold fermions.
Sanner et. al. \cite{Sanner PRL 12} considers two-component fermions with strong repulsive interaction between fermions with different spin projections.
It is demonstrated that pairing instability is faster than the ferromagnetic instability in this regime.
However, the average and weak repulsion shows stable coexisting of fermions with different spin projections.
Interaction of fermions of the same spin projection is not discussed there.
Metastable Stoner-like ferromagnetic phase supported by strong repulsion of fermions
with opposite spin projections in excited scattering states is studied in Ref. \cite{Valtolina Nat Ph 17}.
Mass-imbalanced Fermi-Fermi mixture of $^{161}$Dy and $^{40}$K is created to study the strongly interacting regime  \cite{Ravensbergen PRA 18}.
The existence of the second sound in a system of one-dimensional fermions with repulsive interactions is reported in \cite{Matveev PRL 17},
where hydrodynamic equations for four conserved macroscopic characteristics of the fluid are presented
for the Luttinger liquid with linear excitation spectrum \cite{Matveev PRB 12}, \cite{Luttinger JMP 63}, \cite{Haldane JPC 81}.
This theoretical result corresponds to earlier experimental observation of the first and second sounds in $^{6}$Li atoms \cite{Sidorenkov Nat 13}.

Single state ultracold fermions are also currently studied.
Single spin state degenerate fermions $^{6}$Li confined
by the heavy bosonic atoms $^{133}$Cs at attractive interspecies interaction is created \cite{DeSalvo PRL 17}.
Rakshit et. al. \cite{Rakshit SPP 19} demonstrates that
higher order corrections to the standard mean-field energy can lead
to a formation of Bose-Fermi liquid droplets for attractive BEC and spin-polarized fermions.
A hydrodynamic approach including set of two nonlinear Schrodinger equations is used in \cite{Rakshit SPP 19}.
Weakly interacting single-component two-dimensional dipolar Fermi gas is considered in Ref. \cite{Lu PRA 13} to study the zero sound.

Mixtures of two BECs are studied either.
Bright-bright, dark-bright and dark-dark types of solitons are theoretically obtained in two-component BEC \cite{Csire PRA 10}, \cite{Hoefer PRA 11}.

A possibility that the boson-fermion interaction in the third order by the interaction radius approximation leads
to formation of new soliton is discussed in Ref. \cite{Zezyulin EPJ D 13},
where it is corresponds to a step beyond the mean-field approximation.
However, here we report an advanced version of hydrodynamic of fermions.
Therefore, result of Ref. \cite{Zezyulin EPJ D 13} can be readdressed in terms of novel model,
where less assumptions are made for the force field in the Euler equation.
Moreover, the pressure evolution equation is included here.
The presence of the pressure evolution equation becomes especially important since the force field contains the pressure tensor.
Furthermore, the second interaction constants for each type of interatomic potential are not represented via the scattering length,
but they are considered as independent constants as it follows from their definitions.

Soliton dynamics beyond the mean-field approximation is studied in Refs.
\cite{Katsimiga NJP 17 01}, \cite{Katsimiga NJP 17 02}, \cite{Katsimiga PRA 18}, \cite{Mistakidis NJP 18}.
It is the dark solitons in Bose–Einstein condensates \cite{Katsimiga NJP 17 02}
and bright-dark solitons in two component quantum gases \cite{Katsimiga NJP 17 01}.
Many-body quantum dynamics including the correlation dynamics is considered there.

Microscopic derivation of the quantum hydrodynamic model for boson-fermion mixture, where fermions are located in the single spin state is presented.
General structure of equations for balance of the particle number, momentum and the momentum flux is obtained.
The two-particle short-range boson-boson, boson-fermion, and fermion-fermion interactions are included in the model.
General form of the interaction in the momentum balance equations and the momentum flux evolution equations are found.
Weakly interacting limit of the interaction terms for bosons being in the BEC state and the degenerate fermions is derived up to the third order by the interaction radius.
In this regime the mixture is characterized by five interaction constants.
Small amplitude nonlinear evolution of collective excitations is considered
to find bright-bright and bright-dark solitons in the boson-fermion mixture.
These solitons exist purely due to the interaction constants existing in the third order by the interaction radius.
These solitons are solitons of concentrations of bosons and fermions.
The velocity field of bosons and fermions together with the diagonal elements of pressure reveal the soliton structure either.

This paper is organized as follows.
In Sec. II major steps of derivation of hydrodynamic equations from the Schrodinger equation are demonstrated.
In Sec. III the quantum hydrodynamic equations for BEC and ultracold spin-polarized fermions.
Both species are described up to the third order by the interaction radius is presented.
In Sec IV method of the approximate nonlinear solution is presented.
In Sec. V numerical analysis of obtained Korteweg-–de Vries equation is shown.
In Sec. VI the obtained results are summarized.

\section{On derivation of hydrodynamic equations from microscopic quantum dynamics}

Boson-fermion mixture consisting of $N$ particles,
which is superposition of number of bosons $N_{b}$ and number of fermions $N_{f}$,
is described by the Schrodinger equation with the following Hamiltonian
\begin{equation}\label{BLS20 Hamiltonian micro}
\hat{H}=\sum_{i=1}^{N}\biggl(\frac{\hat{\textbf{p}}^{2}_{i}}{2m_{i}}+V_{ext}(\textbf{r}_{i},t)\biggr)
+\frac{1}{2}\sum_{i,j\neq i}U(\textbf{r}_{i}-\textbf{r}_{j}) ,\end{equation}
where $m_{i}$ is the mass of i-th particle,
$\hat{\textbf{p}}_{i}=-\imath\hbar\nabla_{i}$ is the momentum of i-th particle.
The last term in the Hamiltonian (\ref{BLS20 Hamiltonian micro})
is the short representation of
boson-boson
$(1/2)\sum_{i=1}^{N_{b}} \sum_{j=1,j\neq i}^{N_{b}}U_{bb}(\textbf{r}_{i}-\textbf{r}_{j})$
interparticle interaction,
fermion-fermion
$(1/2)\sum_{i=N_{f}}^{N_{b}+N_{f}} \sum_{j=N_{f},j\neq i}^{N_{b}+N_{f}}U_{ff}(\textbf{r}_{i}-\textbf{r}_{j})$
interparticle interaction,
and boson-fermion
$(1/2)[\sum_{i=1}^{N_{b}} \sum_{j=N_{f}}^{N_{b}+N_{f}}$
$+\sum_{i=N_{f}}^{N_{b}+N_{f}} \sum_{j=1}^{N_{b}}]U_{bf}(\textbf{r}_{i}-\textbf{r}_{j})$
$=\sum_{i=1}^{N_{b}} \sum_{j=N_{f}}^{N_{b}+N_{f}}U_{bf}(\textbf{r}_{i}-\textbf{r}_{j})$
interparticle interaction.
The Schrodinger equation
$\imath\hbar\partial_{t}\Psi(R,t)=\hat{H}\Psi(R,t)$
with Hamiltonian (\ref{BLS20 Hamiltonian micro})
describes the evolution of wave function of full boson-fermion mixture $\Psi(R,t)$,
where full configurational space $R=\{R_{b},R_{f}\}$
is the combination of configurational space of bosons $R_{b}$ and configurational space of fermions $R_{f}$.

However, there are formulations of the many-body problem for fermions with no interaction between fermions in the same spin state
see for instance equation 23 in Ref. \cite{Giorgini RMP 08}.
It is due to the argument that interactions are strongly inhibited by the Pauli exclusion principle.
The antisymmetry of the many-particle wave function is the manifistation of the Pauli exclusion principle.
However, we are not used properties of the wave function at this step.
Mentioning here some of our results we point out that
the first order on the interaction radius (an analog of s-wave) contribution is equal to zero due to the antisymmetry of the many-particle wave function.
However, the third order on the interaction radius (an analog of p-wave) terms are non zero.

At this stage bosons and fermions have arbitrary distributions on quantum states.
Transition to near equilibrium states with zero temperature is made at later stage of derivation,
where truncation of the chain of equations is made.

Concentration of bosons is defined as the quantum mechanical average of the operator of concentration
which is the superposition of the delta functions \cite{MaksimovTMP 2001}, \cite{Andreev PRA08}:
\begin{equation}\label{BLS20 concentration def b} n_{b}(\textbf{r},t)=\int
dR\sum_{i=1}^{N_{b}}\delta(\textbf{r}-\textbf{r}_{i})\Psi^{*}(R,t)\Psi(R,t),\end{equation}
where $dR=dR_{b} dR_{f}$,
$dR_{b}=\prod_{i=1}^{N_{b}}d\textbf{r}_{i}$ is the element of volume in $3N_{b}$ dimensional configurational space,
with $N_{b}$ is the number of bosons, and
$dR_{f}=\prod_{i=N_{b}+1}^{N_{b}+N_{f}}d\textbf{r}_{i}$ is the element of volume in $3N_{f}$ dimensional configurational space,
with $N_{f}$ is the number of fermions.
We need to integrate over coordinates of all particles
since the wave function describes both species.

Definition of the concentration of fermion has similar structure
\begin{equation}\label{BLS20 concentration def f} n_{f}(\textbf{r},t)=\int
dR\sum_{i=N_{b}+1}^{N_{b}+N_{f}}\delta(\textbf{r}-\textbf{r}_{i})\Psi^{*}(R,t)\Psi(R,t),\end{equation}
but operator of concentration of fermions contains coordinates of different set of particles.

Considering time evolution of each concentration $n_{a}$ via the evolution of wave function $\Psi(R,t)$
find the continuity equation \cite{MaksimovTMP 2001}, \cite{Andreev PRA08}:
\begin{equation}\label{BLS20 cont eq via j} \partial_{t}n_{a}+\nabla\cdot \textbf{j}_{a}=0, \end{equation}
where subindex $a$ stands for $b$ or $f$,
and the current $\textbf{j}_{a}$ is defined via the many-particle wave function of the system:
$$\textbf{j}_{a}(\textbf{r},t)
=\int dR\sum_{i\in N_{a}}\delta(\textbf{r}-\textbf{r}_{i})\times$$
\begin{equation}\label{BLS20 j def}
\times\frac{1}{2m_{i}}(\Psi^{*}(R,t)\hat{\textbf{p}}_{i}\Psi(R,t)+c.c.),\end{equation}
where $c.c.$ is the complex conjugation.

Next derive equation for the current evolution.
Consider the time derivative of the current (\ref{BLS20 j def}) using the Schrodinger equation and some straightforward calculations.
As the result find the current evolution equation (it can be also called the momentum evolution equation)
\begin{equation} \label{BLS20 Euler eq 1 via j}
\partial_{t}j_{a}^{\alpha}+\partial_{\beta}\Pi_{a}^{\alpha\beta}
=-\frac{1}{m_{a}}n_{a}\partial_{\alpha}V_{ext}+\frac{1}{m_{a}}F^{\alpha}_{a,int}, \end{equation}
where
$$\Pi_{a}^{\alpha\beta}=\int dR\sum_{i\in N_{a}}\delta(\textbf{r}-\textbf{r}_{i}) \frac{1}{4m_{i}^{2}}
[\Psi^{*}(R,t)\hat{p}_{i}^{\alpha}\hat{p}_{i}^{\beta}\Psi(R,t)$$
\begin{equation} \label{BLS20 Pi def} +\hat{p}_{i}^{\alpha *}\Psi^{*}(R,t)\hat{p}_{i}^{\beta}\Psi(R,t)+c.c.] \end{equation}
is the momentum flux (containing the pressure tensor),
and
\begin{equation} \label{BLS20 F alpha def via n2}
F^{\alpha}_{a,int}=-\sum_{a'=b,f}\int (\partial^{\alpha}U_{aa'}(\textbf{r}-\textbf{r}'))
n_{2,aa'}(\textbf{r},\textbf{r}',t)d\textbf{r}', \end{equation}
with the following expression for the two-particle concentration
$$n_{2,aa'}(\textbf{r},\textbf{r}',t)$$
\begin{equation} \label{BLS20 n2 def} =\int
dR\sum_{i\in N_{a},j\in N_{a'},j\neq i}\delta(\textbf{r}-\textbf{r}_{i})\delta(\textbf{r}'-\textbf{r}_{j})\Psi^{*}(R,t)\Psi(R,t) .\end{equation}

Suggested model includes the pressure evolution equation for fermions.
It is not required for bosons since bosons are considered below in the Bose-Einstein state.
Therefore, the pressure of bosons equals to zero and the bosons are completely described by the concentration and the velocity field.
Therefore, it is required to derive equation for the momentum flux evolution,
since the momentum flux has clear relation to the wave function (\ref{BLS20 Pi def})
similarly to the concentrations and the currents (\ref{BLS20 j def}).
The pressure evolution will be extracted from momentum flux evolution below.
Similarly to derivation of the current evolution equation
consider the time derivative of the momentum flux (\ref{BLS20 Pi def}) using the Schrodinger equation \cite{Andreev 2001}:
$$\partial_{t}\Pi_{f}^{\alpha\beta}+\partial_{\gamma}M_{f}^{\alpha\beta\gamma}
=-\frac{1}{m_{f}}j_{f}^{\beta}\partial_{\alpha}V_{ext}-\frac{1}{m_{f}}j_{f}^{\alpha}\partial_{\beta}V_{ext}$$
$$-\frac{1}{m_{f}}\int[\partial^{\beta}U(\textbf{r}-\textbf{r}')]j_{2}^{\alpha}(\textbf{r},\textbf{r}',t)d\textbf{r}'$$
\begin{equation} \label{BLS20 eq for Pi alpha beta}
-\frac{1}{m_{f}}\int[\partial^{\alpha}U(\textbf{r}-\textbf{r}')]j_{2}^{\beta}(\textbf{r},\textbf{r}',t)d\textbf{r}', \end{equation}
where
$$M_{f}^{\alpha\beta\gamma}=\int dR\sum_{i\in N_{f}}\delta(\textbf{r}-\textbf{r}_{i}) \frac{1}{8m_{i}^{3}}\biggl[\Psi^{*}(R,t)\hat{p}_{i}^{\alpha}\hat{p}_{i}^{\beta}\hat{p}_{i}^{\gamma}\Psi(R,t)$$
$$+\hat{p}_{i}^{\alpha *}\Psi^{*}(R,t)\hat{p}_{i}^{\beta}\hat{p}_{i}^{\gamma}\Psi(R,t)
+\hat{p}_{i}^{\alpha *}\hat{p}_{i}^{\gamma *}\Psi^{*}(R,t)\hat{p}_{i}^{\beta}\Psi(R,t)$$
\begin{equation} \label{BLS20 M alpha beta gamma def}
+\hat{p}_{i}^{\gamma *}\Psi^{*}(R,t)\hat{p}_{i}^{\alpha}\hat{p}_{i}^{\beta}\Psi(R,t)+c.c.\biggr], \end{equation}
and
$$\textbf{j}_{2}(\textbf{r},\textbf{r}',t)=\int
dR\sum_{i\in N_{f}, j\in N, j\neq i}\delta(\textbf{r}-\textbf{r}_{i})\delta(\textbf{r}'-\textbf{r}_{j})\times$$
\begin{equation} \label{BLS20 j 2 def}
\times\frac{1}{2m_{i}}(\Psi^{*}(R,t)\hat{\textbf{p}}_{i}\Psi(R,t)+c.c.) .\end{equation}
If quantum correlations are dropped function $j_{2}^{\alpha}(\textbf{r},\textbf{r}',t)$ splits on product of the current $j_{f}^{\alpha}(\textbf{r},t)$ and the concentration $n_{f}(\textbf{r}',t)$.

Equations (\ref{BLS20 cont eq via j}), (\ref{BLS20 Euler eq 1 via j}), (\ref{BLS20 eq for Pi alpha beta})
are fundamental equations for collection of bosons and fermions.
These equations contain a number of new functions which should be expressed via the basic hydrodynamic functions.
The truncation is to be made for the bosons being in the BEC state and fermions at zero temperature, but collected in the single spin state.

Moreover, it is necessary to present hydrodynamic equations
(\ref{BLS20 cont eq via j}), (\ref{BLS20 Euler eq 1 via j}), (\ref{BLS20 eq for Pi alpha beta}) in more traditional form.
To this end, introduce the velocity field $\textbf{v}_{a}=\textbf{j}_{a}/n_{a}$.
This definition allows to represent the continuity equation in the traditional form.
However, other equations require more detail description which can be found in details in Refs. \cite{Andreev PRA08}, \cite{Andreev 2001}.
the method of introduction of the velocity field includes
the analysis of deviation of velocities of quantum particles introduced as gradients $\hbar\nabla_{i}S/m$ of the phase of wave function $\Psi(R,t)=a(R,t)\exp(\imath S(R,t))$ from the velocity field.
This deviations are also includes the thermal effects and other mechanisms
(Fermi surface caused by the Pauli blocking)
of distribution of particles on quantum states with velocities shifted from the average velocity $\textbf{v}_{a}$.
This method provides the structure of the momentum flux tensor:
\begin{equation} \label{BLS20 Pi via nvv p T}\Pi_{a}^{\alpha\beta}=n_{a}v_{a}^{\alpha}v_{a}^{\beta} +p_{a}^{\alpha\beta}+T_{a}^{\alpha\beta}.\end{equation}
The first and second terms on the right-hand side of equation (\ref{BLS20 Pi via nvv p T})
have classical meaning and include the pressure tensor $p^{\alpha\beta}$.
The last term in equation (\ref{BLS20 Pi via nvv p T}) has quantum nature and can be presented in the following approximate form
\begin{equation} \label{BLS20 Bohm tensor single part}
T_{a}^{\alpha\beta}=-\frac{\hbar^{2}}{4m_{a}^{2}}\biggl[\partial_{\alpha}\partial_{\beta}n_{a}
-\frac{\partial_{\alpha}n_{a}\cdot\partial_{\beta}n_{a}}{n_{a}}\biggr].\end{equation}
It is related to the quantum Bohm potential.

Equation (\ref{BLS20 Bohm tensor single part}) appears for noninteracting bosons in the BEC state.
Its linear part (the first term) is straightforward for interacting bosons or interacting fermions.
While the second term in equation (\ref{BLS20 Bohm tensor single part}) has no proper justification for fermions even if interaction is neglected.
Hence, the second term in equation (\ref{BLS20 Bohm tensor single part}) for fermions is an approximate equation of state.

Representation for tensor $M_{f}^{\alpha\beta\gamma}$ (\ref{BLS20 M alpha beta gamma def}) similar to (\ref{BLS20 Bohm tensor single part})
can be found either.
It is given in Ref. \cite{Andreev 2001} as set of four rather large equations (see equations 25-28).

Equations are obtained for particles with arbitrary spin,
but further analysis is made for spin-0 bosons and spin-1/2 fermions being in the single spin projection state.

Equations (\ref{BLS20 cont eq via j})-(\ref{BLS20 eq for Pi alpha beta}) are derived for arbitrary potential $U_{ij}=U(\textbf{r}_{i}-\textbf{r}_{j})$.
It is necessary to specify that neutral particles interact via the short-range potential.
To stress the small radius of interaction represent the coordinates of interacting particles $\textbf{r}_{i}$ and $\textbf{r}_{j}$
via the relative distance and coordinate of their center of mass.
Next, we can expand the delta functions and the wave function on the small interparticle distance
$\textbf{r}_{ij}=\textbf{r}_{i}-\textbf{r}_{j}$
since potential $U_{ij}$ goes to zero at the large interparticle distances.
The straightforward calculations for weakly interacting particles including symmetry between fermions,
between bosons, and the absence of symmetry between bosons and fermions (for more details see \cite{Andreev PRA08}, \cite{Andreev 2001}).
Terms in the zeroth order on the interparticle distance cancel each other.
In the first order, there are nonzero terms for boson-boson and boson-fermion interactions,
which corresponds to the Gross-Pitaevskii approximation. 
The integral over interparticle distance contains the potential of interaction and gives the interaction constant for each interaction.
The fermion-fermion interaction term in the first order equals to zero due to the antisymmetry of the wave function.
Formally, we have the first interaction constant for fermions, but it is multiplied by the function which equal to zero.
The second order terms go to zero due to integration over the interparticle distance (its angular dependence).
The third order of expansion gives nonzero results for all three interactions.
The second interaction constant appears for each interaction.

The derivation of fundamental hydrodynamic equations is made by the many-particle quantum hydrodynamic method
\cite{Andreev LP 19}, \cite{MaksimovTMP 2001}, \cite{Andreev PRA08}.
Paying attention to development of hydrodynamic methods mention the generalized hydrodynamics actively developing in recent years
\cite{Ruggiero arxiv 19}, \cite{Bertini PRL 16}.

\section{Hydrodynamic equations for boson-fermion mixture}

In this regime we have two continuity equations:
\begin{equation}\label{BLS20 cont eq bosons} \partial_{t}n_{b}+\nabla\cdot (n_{b}\textbf{v}_{b})=0, \end{equation}
and
\begin{equation}\label{BLS20 cont eq fermions} \partial_{t}n_{f}+\nabla\cdot (n_{f}\textbf{v}_{f})=0. \end{equation}

We also have two Euler (momentum balance) equations.
The Euler equation for bosons
$$m_{b}n_{b}(\partial_{t} +\textbf{v}_{b}\cdot\nabla)v^{\alpha}_{b}
-\frac{\hbar^{2}}{2m_{b}}n_{b}\partial^{\alpha}\frac{\triangle\sqrt{n_{b}}}{\sqrt{n_{b}}} $$
$$+g_{b} n_{b}\partial^{\alpha}n_{b} +\frac{1}{2}g_{2b} \partial^{\alpha}\triangle n_{b}^{2} =-n_{b}\partial^{\alpha}V_{ext}$$
\begin{equation}\label{BLS20 Euler bosons}
-g_{bf} n_{b}\partial^{\alpha}n_{f}-\frac{g_{2,bf}}{2}n_{b}\partial^{\alpha}\triangle n_{f}\end{equation}
contains the boson-boson interaction in the first and third orders by the interaction radius,
the terms proportional to $g_{b}$ and $g_{2b}$ constants, correspondingly.
It includes the boson-fermion interaction
which are proportional to $g_{bf}$ in the first order and $g_{2bf}$ in the third order.

The Euler equation for fermions is
$$m_{f}n_{f}(\partial_{t} +\textbf{v}_{f}\cdot\nabla)v^{\alpha}_{f}
-\frac{\hbar^{2}}{2m_{f}}n_{f}\partial^{\alpha}\frac{\triangle\sqrt{n_{f}}}{\sqrt{n_{f}}}+\partial^{\beta}p_{f}^{\alpha\beta} $$
\begin{equation}\label{BLS20 Euler fermions}
=-g_{bf} n_{f}\partial^{\alpha}n_{b}-\frac{g_{2,bf}}{2}n_{f}\partial^{\alpha}\triangle n_{b} -g_{2f}\frac{m_{f}^{2}}{2\hbar^{2}}I_{0}^{\alpha\beta\gamma\delta}\partial^{\beta}(n_{f}p^{\gamma\delta}_{f}),\end{equation}
where all fermions are in the quantum states with the same spin projection.
Euler equation (\ref{BLS20 Euler fermions}) has contribution of the boson-fermion interaction.
Similar to the Euler equation for bosons (\ref{BLS20 Euler bosons}),
they are proportional to $g_{bf}$ in the first order by the interaction radius and $g_{2bf}$ in the third order.
The fermion-fermion interaction gives the single term in Euler equation (\ref{BLS20 Euler fermions}).
It appears in the third order by the interaction radius being proportional to $g_{2f}$ constant.

The third order by the interaction radius approximation has similarity to the p-wave interaction.
The p-wave fermion-fermion interaction is studied in the boson-fermion mixtures at study of solitons in mixtures \cite{Adhikari JPB05}.
Traditional p-wave approximation assumes an equation of state for the pressure in terms of the concentration.
However, our model gives more accurate analysis of pressure via the pressure evolution equation.

The boson-boson interaction in the third order by the interaction radius is presented by the nonlocal interaction term showing similarity to the models presented in Refs. \cite{Rosanov}, \cite{Braaten}.

The pressure evolution equation for fermions is also a part of developed and applied hydrodynamic model
$$\partial_{t}p_{f}^{\alpha\beta} +v_{f}^{\gamma}\partial_{\gamma}p_{f}^{\alpha\beta} +p_{f}^{\alpha\gamma}\partial_{\gamma}v_{f}^{\beta} +p_{f}^{\beta\gamma}\partial_{\gamma}v_{f}^{\alpha}
+p_{f}^{\alpha\beta}\partial_{\gamma}v_{f}^{\gamma} $$
$$=-\frac{m_{f}}{8\hbar^{2}}g_{2f} \{I_{0}^{\alpha\gamma\delta\mu}
[3 n_{f}^{2} v_{f}^{\beta}v_{f}^{\delta}\partial^{\gamma}v_{f}^{\mu}
+2n_{f}p_{f}^{\mu\delta}(\partial^{\gamma}v_{f}^{\beta}-\partial^{\beta}v_{f}^{\gamma})]$$
\begin{equation} \label{BLS20 pressure evolution}
+I_{0}^{\beta\gamma\delta\mu}
[3 n_{f}^{2} v_{f}^{\alpha}v_{f}^{\delta}\partial^{\gamma}v_{f}^{\mu}
+2n_{f}p_{f}^{\mu\delta}(\partial^{\gamma}v_{f}^{\alpha}-\partial^{\alpha}v_{f}^{\gamma})]\},  \end{equation}
where
$I_{0}^{\alpha\beta\gamma\delta}=
\delta^{\alpha\beta}\delta^{\gamma\delta}+\delta^{\alpha\gamma}\delta^{\beta\delta}+\delta^{\alpha\delta}\delta^{\beta\gamma}$.

Equations (\ref{BLS20 cont eq bosons})-(\ref{BLS20 pressure evolution}) contain the following interaction constants
$g_{b}=\int U_{bb}d\textbf{r} $,
$g_{bf}=\int U_{bf}d\textbf{r} $,
$g_{2b}=(1/24)\int r^{2}U_{bb}d\textbf{r} $,
$g_{2bf}=\int r^{2}U_{bf}d\textbf{r} $,
and
$g_{2f}=\int r^{2}U_{ff}d\textbf{r} $.

The pressure evolution equation (\ref{BLS20 pressure evolution}) appears from equation (\ref{BLS20 eq for Pi alpha beta})
after extraction of the thermal part or other mechanisms of distribution of particles in the momentum space like the Pauli blocking for degenerate fermions.

Equation (\ref{BLS20 pressure evolution}) has no trace of the external potential and boson-fermion interaction.

The fermion-fermion interaction gives nonzero contribution in the third order by the interaction radius.
It consists of the structure of two terms
which is repeated twice to give it the form symmetric on free indexes
since the pressure tensor $p_{f}^{\alpha\beta}$ is a symmetric tensor.
Let us have a closer look on each of two terms.
One is highly nonlinear and includes product of three velocities $n_{f}^{2} v_{f}^{\beta}v_{f}^{\delta}\partial^{\gamma}v_{f}^{\mu}$.
Another term is proportional to the pressure tensor.
Moreover, it is proportional to the classical hydrodynamic vorticity
$\varepsilon^{\gamma\beta\delta}\Omega^{\delta}$
$=\partial^{\gamma}v_{f}^{\beta}-\partial^{\beta}v_{f}^{\gamma}$,
where
$\Omega^{\alpha}=\varepsilon^{\alpha\beta\gamma}\partial_{\beta}v_{\gamma}$ is the vorticity of classic uncharged fluid.

The left-hand side of equation (\ref{BLS20 pressure evolution}) contains
the divergence of a third rank tensor $\partial_{\gamma}Q^{\alpha\beta\gamma}$
which is the average of product of three thermal velocities
(the thermal part of tensor $M_{f}^{\alpha\beta\gamma}$ (\ref{BLS20 M alpha beta gamma def})),
while $p^{\alpha\beta}$ is the average of product of two thermal velocities.
It is assumed to be equal to zero.
It is the equation of state obtained as extension of equilibrium value of tensor $Q^{\alpha\beta\gamma}$.

The boson part of the model is developed in Refs. \cite{Andreev LP 19}, \cite{Andreev PRA08}.
The fermion part of the hydrodynamic model is derived in Refs. \cite{Andreev 2001}, \cite{Andreev 1912}.
Here, same is in Ref. \cite{Andreev 1912}, the pressure evolution equation is considered in the long-wavelength limit,
so high order derivatives are neglected.
The interspecies interaction is addressed in terms of many-particle quantum hydrodynamic method in Refs. \cite{Andreev PRA08}, \cite{Andreev 2001}.

\section{Perturbation method}

Following papers \cite{Andreev MPL B 12} and \cite{Andreev LP 19}
we use the reductive perturbation method \cite{Washimi PRL 66}, \cite{Kalita PlasmaPhys 98}
to study the.
According to this method all hydrodynamic values may be
represented as:

\begin{equation}\label{BLS20 reduction of n b}
n_b=n_{0b}+\varepsilon n_{1b}+\varepsilon^2 n_{2b}+... ,\end{equation}
\begin{equation}\label{BLS20 reduction of n f}
n_f=n_{0f}+\varepsilon n_{1f}+\varepsilon^2 n_{2f}+... ,\end{equation}
\begin{equation}\label{BLS20 reduction of v b}
v_{b}^{x}=\varepsilon v_{1b}+\varepsilon^2 v_{2b}+... ,\end{equation}
\begin{equation}\label{BLS20 reduction of v f}
v_{f}^{x}=\varepsilon v_{1f}+\varepsilon^2 v_{2f}+... ,\end{equation}
and
\begin{equation}\label{BLS20 reduction of p f}
p_{f}^{ii}=p_{0f}^{ii}+\varepsilon p_{1f}^{ii}+\varepsilon^2 p_{2f}^{ii}+... ,\end{equation}
where $ii$ stands for $xx$, $yy$ and $zz$
since all diagonal elements of the pressure tensor are involved in dynamics of longitudinal perturbations.

It is assumed that there are nonzero constant equilibrium concentrations and pressure of fermions.
The velocity fields are equal to zero in equilibrium.

We also performed the following "scaling" of variables:
\begin{equation}\label{BLS20 scaling of x}
\xi = \varepsilon ^{1/2}(x-Vt)
\end{equation}
and
\begin{equation}\label{BLS20 scaling of t}
\tau = \varepsilon ^{3/2}Vt .
\end{equation}
The latter expression introduces so-called "slow" time.

\subsection{The first order perturbations}

Substitute scaling of hydrodynamic functions (\ref{BLS20 reduction of n b})-(\ref{BLS20 reduction of p f})
and space-time variables (\ref{BLS20 scaling of x}), (\ref{BLS20 scaling of t})
in basic equations (\ref{BLS20 cont eq bosons})-(\ref{BLS20 pressure evolution}).
Separate contributions appearing in different orders on parameter $\varepsilon$.
Extract equations in the lowest order on parameter $\varepsilon$ and find
the continuity equation for bosons
\begin{equation} \label{BLS20 cont b I order}n_{0b}\partial_{\xi}v_{1b}-V\partial_{\xi}n_{1b}=0,\end{equation}
the Euler equation for bosons
\begin{equation} \label{BLS20 Euler b I order}m_{b}V\partial_{\xi}v_{1b}-g_{b}\partial_{\xi}n_{1b}-g_{bf}\partial_{\xi}n_{1f}=0,\end{equation}
the continuity equation for fermions
\begin{equation} \label{BLS20 cont f I order}n_{0f}\partial_{\xi}v_{1f}-V\partial_{\xi}n_{1f}=0,\end{equation}
the Euler equation for fermions
$$m_{f}n_{0f}V\partial_{\xi}v_{1f}+\partial_{\xi}p_{1f}^{xx}=-g_{bf}n_{0f}\partial_{\xi}n_{1b}$$
\begin{equation} \label{BLS20 Euler f I order}-g_{2f}\frac{m_{f}^{2}}{2\hbar^{2}}
\partial_{\xi}[n_{0f}(3p_{1f}^{xx}+p_{1f}^{yy}+p_{1f}^{zz})+(3p_{0f}^{xx}+p_{0f}^{yy}+p_{0f}^{zz})n_{1f}],\end{equation}
and equations for evolution of the elements of pressure tensor
\begin{equation} \label{BLS20 pressure xx f I order} V\partial_{\xi}p_{1f}^{xx}-3p_{0f}^{xx}\partial_{\xi}v_{1f}=0,\end{equation}
and
\begin{equation} \label{BLS20 pressure yy zz f I order}V\partial_{\xi}p_{1f}^{yy}-p_{0f}^{yy}\partial_{\xi}v_{1f}=0,\end{equation}
$zz$ element is the same as $yy$ element.

Fermi surface in equilibrium regime is assumed to be a sphere.
Therefore, we have $p_{0f}^{xx}=p_{0f}^{yy}=p_{0f}^{zz}\equiv p_{0f}$.

Equations (\ref{BLS20 pressure xx f I order})-(\ref{BLS20 pressure yy zz f I order}) show
that the perturbation of pressure in the direction of wave propagation is three times larger
than the perturbation in the perpendicular directions.

\begin{figure}
\includegraphics[width=8cm,angle=0]{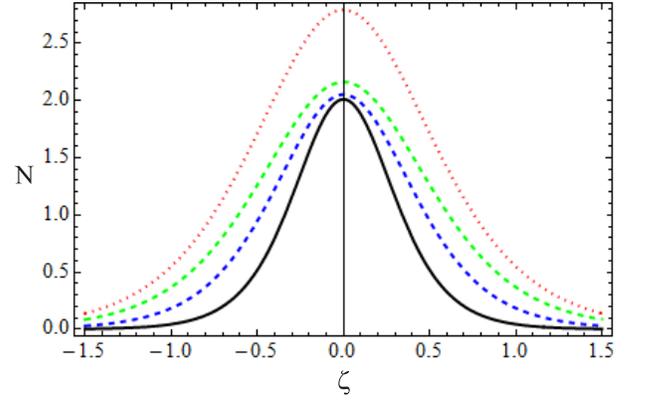}
\caption{\label{BLS20 Fig form of n b 1}
Form of soliton obtained for concentration of bosons $N\equiv n_{1b}/n_{0b}$
as function of $\zeta=\tau-U\xi$ given by equation (\ref{BLS20 solution for n 1 b}).
Solution is demonstrated at different boson-boson interaction constant $g$.
The upper red dotted line corresponds to $g=1$.
The second from above line (the green dashed line) corresponds to $g=3$.
The third from above line (the blue dashed line) corresponds to $g=5$.
The lowest black solid line corresponds to $g=10$.
Other parameters are kept at the following values: $m=4$, $n=2$, $l=1$, $L=0$, $F=0$, $G=1$.
Change of $g$ suggest some change of $G$
which is neglected.
Value of $G$ is kept well above critical value $G_{min}=0.25$.}
\end{figure}

\begin{figure}
\includegraphics[width=8cm,angle=0]{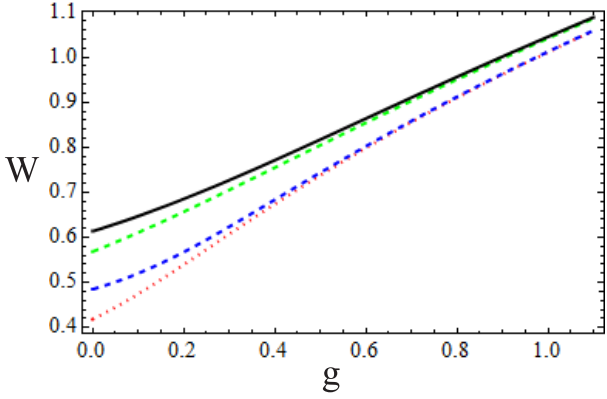}
\caption{\label{BLS20 Fig W 1}
Dimensionless perturbation velocity (\ref{BLS20 W-2}) is plotted
for different combinations of boson-fermion interaction $l$ and
fermion-fermion interaction in the third order $F$
as the function of boson-boson interaction $g$.
The mass ratio $M_{0}=9$ and the concentration ratio $N_{0}=0.2$ are fixed.
Parameters $G$ and $L$ give no contribution into velocity (\ref{BLS20 W-2}).
%Thick, Red, Dotted; Thick,
The lower red dotted line corresponds to $l = 1$ and $F = 0.1$.
The upper black solid line corresponds to $l = 2$ and $F = 1$.
The second from above line at large $g$ (the green light-dashed line) corresponds to $l = 2$ and $F = 0.1$.
The third from above line at large $g$ (the blue dark-dashed line) corresponds to $l = 1$ and $F = 1$.
}
\end{figure}

Equations (\ref{BLS20 cont b I order})-(\ref{BLS20 pressure yy zz f I order}) are uniform linear differential equations.
We obtain explicit relation between these functions using boundary conditions at infinity,
where all perturbations tend to zero.
They have nonzero solution if parameter $V$ have the following form
$$V^{2}=\frac{1}{2}\biggl[\frac{n_{0b}g_{b}}{m_{b}}+\frac{3p_{0f}}{m_{f}n_{0f}}+g_{2f}\frac{8p_{0f}m_{f}}{\hbar^{2}}$$
\begin{equation} \label{BLS20 solution for V} \pm\sqrt{\biggl(\frac{n_{0b}g_{b}}{m_{b}}-\frac{3p_{0f}}{m_{f}n_{0f}}-g_{2f}\frac{8p_{0f}m_{f}}{\hbar^{2}}\biggr)^{2}
+4\frac{n_{0b}n_{0f}g_{bf}^{2}}{m_{b}m_{f}}}\biggr],\end{equation}
"+" corresponds to perturbations in system of bosons affected by fermions,
"-" corresponds to perturbations in system of fermions affected by bosons.

Consider sign "+" in front of the square root and drop contribution of fermions
then we get
\begin{equation} \label{BLS20 V b}V_{b}^{2}=\frac{n_{0b}g_{b}}{m_{b}}\end{equation}
velocity for nonlinear perturbations in BEC considered in Refs. \cite{Andreev MPL B 12}, \cite{Andreev LP 19}.
This velocity corresponds to the long-wavelength limit of the Bogoliubov spectrum.
The velocity square $V^{2}$ is positive for the repulsive interaction between bosons.
Consider the influence of fermions on the boson solution.
If boson-fermion interaction is small we can expand the square root.
After expansion assuming that the partial velocity of bosons (\ref{BLS20 V b}) dominates
over the partial velocity of fermions $V_{f}^{2}=3p_{0f}/m_{f}n_{0f}+8 g_{2f}p_{0f}m_{f}/\hbar^{2}$
we find the following expression
\begin{equation} \label{BLS20 solution for V expanded} V^{2}=\frac{n_{0b}g_{b}}{m_{b}}+
\frac{ \frac{n_{0b}n_{0f}g_{bf}^{2}}{m_{b}m_{f}}}{
(\frac{n_{0b}g_{b}}{m_{b}} -\frac{3p_{0f}}{m_{f}n_{0f}} -g_{2f} \frac{8p_{0f}m_{f}}{\hbar^{2}})}.\end{equation}
General behavior of velocity (\ref{BLS20 solution for V}) shows that the chosen solution should have positive second term.
The sign of boson-fermion interaction does not affect the velocity of nonlinear perturbations.
It is corresponds to the general solution (\ref{BLS20 solution for V}).

Solution (\ref{BLS20 solution for V}) with sign $-$ corresponds to the acoustic wave in fully spin polarized fermions.
Separation on the bosonic and fermionic branches is partially conventional.
If the the partial velocity of fermions $V_{f}$ dominates over the partial velocity of bosons $V_{b}$ (\ref{BLS20 V b})
we have sign $-$ for bosonic branch (find solution (\ref{BLS20 V b}) for small boson-fermion interaction) and sign $+$ for fermionic branch.
However, study of the mixture for intermediate boson-fermion interaction does not allow so straightforward separation on bosonic and fermionic branches.
So, we keep studying nonlinear solution corresponding to sign "+" in (\ref{BLS20 solution for V}) and we conventionally call it the bosonic branch.
The second branch conventionally called the fermionic branch will be studied elsewhere.

Fermions in partially polarized regime demonstrate two acoustic wave and the spin wave with $\omega(k=0)\neq0$ \cite{Andreev LPL 18}.
Hamiltonian of the nonlinear Pauli equation in \cite{Andreev LPL 18} contains
the interaction term corresponding to the total energy of a two-component Fermi gas presented in \cite{Jo Sc 09}.

\begin{figure}
\includegraphics[width=8cm,angle=0]{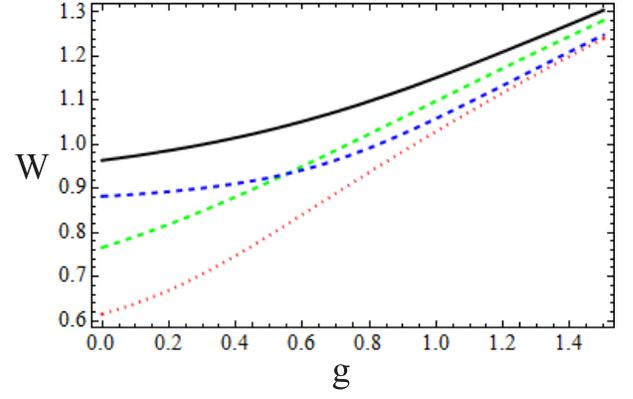}
\caption{\label{BLS20 Fig W 2}  %m = 4  n = 0.2
Dimensionless perturbation velocity (\ref{BLS20 W-2}) is plotted
for different combinations of boson-fermion interaction $l$ and
fermion-fermion interaction in the third order $F$
as the function of boson-boson interaction $g$.
The mass ratio $M_{0}=4$ and the concentration ratio $N_{0}=0.2$ are fixed.
Parameters $G$ and $L$ give no contribution into velocity (\ref{BLS20 W-2}).
The lower red dotted line corresponds to $l = 1$ and $F = 0.1$.
The upper black solid line corresponds to $l = 2$ and $F = 1$.
The second from above line (the green dashed line) corresponds to $l = 2$ and $F = 0.1$.
The third from above line (the blue dashed line) corresponds to $l = 1$ and $F = 1$.}
\end{figure}

\begin{figure}
\includegraphics[width=8cm,angle=0]{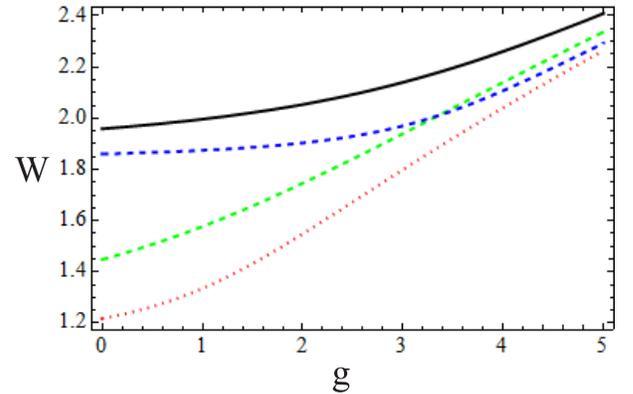}
\caption{\label{BLS20 Fig W 3} %m = 4  n = 2
Dimensionless perturbation velocity (\ref{BLS20 W-2}) is plotted
for different combinations of boson-fermion interaction $l$ and
fermion-fermion interaction in the third order $F$
as the function of boson-boson interaction $g$.
The mass ratio $M_{0}= 4$ and the concentration ratio $N_{0}= 2$ are fixed.
Parameters $G$ and $L$ give no contribution into velocity (\ref{BLS20 W-2}).
The lower red dotted line corresponds to $l = 1$ and $F = 0.1$.
The upper black solid line corresponds to $l = 2$ and $F = 1$.
The second from above line (the green dashed line) corresponds to $l = 2$ and $F = 0.1$.
The third from above line (the blue dashed line) corresponds to $l = 1$ and $F = 1$.}
\end{figure}

\subsection{The second order perturbations}

Next order on $\varepsilon$ contribution in hydrodynamic equations leads to the following set of nonlinear differential equations:
the continuity equation for bosons
\begin{equation} \label{BLS20 cont b II order} V\partial_{\tau}n_{1b}-V\partial_{\xi}n_{2b}+n_{0b}\partial_{\xi}v_{2b}+\partial_{\xi}(n_{1b}v_{1b})=0;\end{equation}
the Euler equation for bosons
$$m_{b}n_{0b}V\partial_{\tau}v_{1b}-m_{b}n_{0b}V\partial_{\xi}v_{2b}-m_{b}n_{1b}V\partial_{\xi}v_{1b}
-\frac{\hbar^{2}}{4m_{b}}\partial_{\xi}^{3}n_{1b}$$
$$+m_{b}n_{0b}v_{1b}V\partial_{\xi}v_{1b}
=-g_{b}n_{0b}\partial_{\xi}n_{2b}-g_{b}n_{1b}\partial_{\xi}n_{1b}
-g_{2b}n_{0b}\partial_{\xi}^{3}n_{1b}$$
\begin{equation} \label{BLS20 Euler b II order} -g_{bf}n_{0b}\partial_{\xi}n_{2f} -g_{bf}n_{1b}\partial_{\xi}n_{1f}
-\frac{1}{2}g_{2bf}n_{0b}\partial_{\xi}^{3}n_{1f};\end{equation}
the continuity equation for fermions
\begin{equation} \label{BLS20 cont f II order} V\partial_{\tau}n_{1f}-V\partial_{\xi}n_{2f}+n_{0f}\partial_{\xi}v_{2f}+\partial_{\xi}(n_{1f}v_{1f})=0;\end{equation}
the Euler equation for fermions
$$m_{f}n_{0f}V\partial_{\tau}v_{1f}-m_{f}n_{0f}V\partial_{\xi}v_{2f}-m_{f}Vn_{1f}\partial_{\xi}v_{1f}$$
$$+m_{f}n_{0f}v_{1f}\partial_{\xi}v_{1f} -\frac{\hbar^{2}}{4m_{f}}\partial_{\xi}^{3}n_{1f}+\partial_{\xi}p_{2}^{xx}$$
$$=-g_{2f}\frac{m^{2}}{2\hbar^{2}}\partial_{\xi}[n_{0f}(3p_{2f}^{xx}+p_{2f}^{yy}+p_{2f}^{zz})$$
$$+(3p_{0f}^{xx}+p_{0f}^{yy}+p_{0f}^{zz})n_{2f} +(3p_{1f}^{xx}+p_{1f}^{yy}+p_{1f}^{zz})n_{1f}]$$
\begin{equation} \label{BLS20 Euler f II order} -g_{bf}n_{0f}\partial_{\xi}n_{2b}-g_{bf}n_{1f}\partial_{\xi}n_{1b}
-\frac{1}{2}g_{2bf}n_{0f}\partial_{\xi}^{3}n_{1b};\end{equation}
equations for evolution of the elements of pressure tensor
\begin{equation} \label{BLS20 pressure xx f II order} V\partial_{\tau}p_{1f}^{xx}-V\partial_{\xi}p_{2f}^{xx}+v_{1f}\partial_{\xi}p_{1f}^{xx}
+3p_{0f}^{xx}\partial_{\xi}v_{2f}+3p_{1f}^{xx}\partial_{\xi}v_{1f}=0,\end{equation}
\begin{equation} \label{BLS20 pressure yy f II order} V\partial_{\tau}p_{1f}^{yy}-V\partial_{\xi}p_{2f}^{yy}+v_{1f}\partial_{\xi}p_{1f}^{yy}
+p_{0f}^{yy}\partial_{\xi}v_{2f}+p_{1f}^{yy}\partial_{\xi}v_{1f}=0,\end{equation}
and
\begin{equation} \label{BLS20 pressure zz f II order} V\partial_{\tau}p_{1f}^{zz}-V\partial_{\xi}p_{2f}^{zz}+v_{1f}\partial_{\xi}p_{1f}^{zz}
+p_{0f}^{zz}\partial_{\xi}v_{2f}+p_{1f}^{zz}\partial_{\xi}v_{1f}=0.\end{equation}
All functions of the first order can be represented
via the first order perturbations for concentrations of bosons $n_{1b}$ and fermions $n_{1f}$.
Presenting the second order hydrodynamic perturbations via the derivative of second order concentration of fermions $\partial_{\xi}n_{2f}$
and first order perturbations for concentrations of bosons $n_{1b}$ and fermions $n_{1f}$
find equation for three variables,
where coefficient in front of $\partial_{\xi}n_{2f}$ equals to zero
if expression (\ref{BLS20 solution for V}) for $V^{2}$ is included.

After described manipulations obtain equation for concentrations $n_{2f}$, $n_{1f}$, $n_{1b}$: \begin{widetext}
$$g_{bf}n_{0f}\cdot g_{bf}n_{0b}\partial_{\xi}n_{2f}
+\partial_{\xi}n_{2f}
\biggl[8g_{2f}p_{0}\frac{m_{f}^{2}}{\hbar^{2}}+\frac{3p_{0}}{n_{0f}}-m_{f}V^{2}\biggr](m_{b}V^{2}-g_{b}n_{0b})$$
$$+(m_{b}V^{2}-g_{b}n_{0b})\biggl[2m_{f}V^{2}\biggl(\partial_{\tau}n_{1f}+\frac{n_{1f}\partial_{\xi}n_{1f}}{n_{0f}}\biggr)
-\frac{\hbar^{2}}{4m_{f}}\partial_{\xi}^{3}n_{1f}
+n_{1f}\partial_{\xi}n_{1f}\biggl(\frac{6p_{0}}{n_{0f}^{2}}+20g_{2f}\frac{p_{0}m_{f}^{2}}{n_{0f}\hbar^{2}}\biggr)$$
$$+g_{bf}n_{1f}\partial_{\xi}n_{1b}+\frac{1}{2}g_{2bf}n_{0f}\partial_{\xi}^{3}n_{1b}\biggr]
+g_{bf}n_{0f}\biggl[2m_{b}V^{2}\partial_{\tau}n_{1b}+2m_{b}V^{2}\frac{n_{1b}\partial_{\xi}n_{1b}}{n_{0b}}-\frac{\hbar^{2}}{4m_{b}}\partial_{\xi}^{3}n_{1b}$$
\begin{equation} \label{BLS20 eq for b and f concentrations} +g_{b}n_{1b}\partial_{\xi}n_{1b}+g_{2b}n_{0b}\partial_{\xi}^{3}n_{1b}
+g_{bf}n_{1b}\partial_{\xi}n_{1f}+\frac{1}{2}g_{2bf}n_{0b}\partial_{\xi}^{3}n_{1f}\biggr]=0.\end{equation}
\end{widetext}
The first and second terms in equation (\ref{BLS20 eq for b and f concentrations}) contain all contribution of the second order functions
(in this case they are expressed via the second order concentration of fermions $n_{2f}$).
%It can be seen that their contribution is equal to zero if we include the explicit form of
The coefficient in front of $n_{2f}$ goes to zero
if explicit form of velocity $V$ (\ref{BLS20 solution for V}) is used.
Therefore, equation (\ref{BLS20 eq for b and f concentrations}) reduces to equation relatively two functions $n_{1b}$ and $n_{1f}$.

The first term in equation (\ref{BLS20 eq for b and f concentrations})
and combined inside square brackets group of seven last terms present the contribution of bosons.
Other terms present the contribution of fermions.

Equation (\ref{BLS20 eq for b and f concentrations}) appears as the Euler equation for fermions.
So the contribution of bosons equal to zero if the interspecies interaction constant $g_{bf}$ goes to zero.

The lowest order on $\varepsilon$ analysis gives relation between concentrations of bosons and fermions
\begin{equation} \label{BLS20 rel between n_f and n_b}
n_{1f}=\frac{m_{b}}{n_{0b}g_{bf}}\biggl(V^{2}-\frac{n_{0b}g_{b}}{m_{b}}\biggr)n_{1b}.\end{equation}
This relation allows to get an equation for single function ($n_{1b}$ for instance) from (\ref{BLS20 eq for b and f concentrations}).
Next, then $n_{1b}$ is found we obtain the structure of soliton solution for $n_{1f}$ using relation (\ref{BLS20 rel between n_f and n_b}).

Moreover, we can use solution for $V^{2}$ given by equation (\ref{BLS20 solution for V}) to analyze relation (\ref{BLS20 rel between n_f and n_b}).
$$\frac{n_{1f}}{n_{1b}}=\frac{1}{2}\frac{m_{b}}{n_{0b}g_{bf}}\Biggl[\frac{3p_{0}}{m_{f}n_{0f}}+g_{2f}\frac{8p_{0}m_{f}}{\hbar^{2}}-\frac{n_{0b}g_{b}}{m_{b}}$$
\begin{equation} \label{BLS20 rel between n_f and n_b explicit} +\sqrt{\biggl(\frac{3p_{0}}{m_{f}n_{0f}}+g_{2f}\frac{8p_{0}m_{f}}{\hbar^{2}}-\frac{n_{0b}g_{b}}{m_{b}}\biggr)^{2}
+4\frac{n_{0b}n_{0f}g_{bf}^{2}}{m_{b}m_{f}}}\Biggr].\end{equation}
The right-hand side of equation (\ref{BLS20 rel between n_f and n_b explicit}) is the product of two functions:
the interaction constant $g_{bf}$
and combination of parameters located in brackets.
The structure of the parameters in brackets can be expressed as follows: $\Xi+\sqrt{\Xi^{2}+\Lambda^{2}}$.
The sign of this structure does not depend on signs and values of parameters $\Xi$, $\Lambda$ being always positive.
Therefore, perturbations for the bosons and fermions have same sign if boson-fermion interaction is repulsive $g_{bf}>0$
or they have opposite signs if boson-fermion interaction is attractive $g_{bf}<0$.

Expression (\ref{BLS20 rel between n_f and n_b explicit}) can be rewritten in different equivalent form
\begin{equation} \label{BLS20 rel between n_f and n_b f2}
n_{1f}=\frac{n_{0f}g_{bf}}{m_{f}} \frac{1}{(V^{2}-\frac{3p_{0f}}{m_{f}n_{0f}}-\frac{8g_{2f}p_{0f}m_{f}}{\hbar^{2}})}n_{1b}.\end{equation}

\subsection{Korteweg-–de Vries equation for perturbations of bosons}

Korteweg-–de Vries (KdV) equation for concentration of bosons has the following structure
\begin{equation} \label{BLS20 KdV} \tilde{a}\partial_{\tau}n_{1b}+\tilde{b}n_{1b}\partial_{\xi}n_{1b}+\tilde{c}\partial_{\xi}^{3}n_{1b}=0,\end{equation}
where we find coefficients
\begin{equation} \label{BLS20 a tilde} \tilde{a} =2m_{b}V^{2}
\Biggl[1 +
\frac{ \frac{n_{0f}g_{bf}}{m_{f}}\frac{n_{0b}g_{bf}}{m_{b}} }{ (V^{2} -\frac{3p_{0}}{m_{f}n_{0f}}-\frac{8p_{0f}g_{2f}m_{f}}{\hbar^{2}})^{2} }\Biggr],\end{equation}

$$\tilde{b}=\frac{1}{n_{0b}} \biggl[ 2m_{b}V^{2} +g_{b}n_{0b} +g_{bf}n_{0b}
\frac{ \frac{n_{0f}g_{bf}}{m_{f}}}{V^{2} -\frac{3p_{0}}{m_{f}n_{0f}}-\frac{8p_{0f}g_{2f}m_{f}}{\hbar^{2}}}$$
$$+g_{bf}n_{0b}
\frac{\frac{n_{0f}g_{bf}}{m_{f}}\frac{n_{0b}g_{bf}}{m_{f}}}{(V^{2} -\frac{3p_{0}}{m_{f}n_{0f}}-\frac{8p_{0f}g_{2f}m_{f}}{\hbar^{2}})^{2}}$$
$$+m_{b}\frac{n_{0f}}{n_{0b}}\biggl(\frac{m_{f}}{m_{b}}\biggr)^{-2}
\biggl(V^{2} +\frac{3p_{0}}{m_{f}n_{0f}} +\frac{10p_{0f}g_{2f}m_{f}}{\hbar^{2}}\biggr)
\times$$
\begin{equation} \label{BLS20 b tilde} \times
\frac{(\frac{n_{0b}g_{bf}}{m_{b}})^{3}}{(V^{2} -\frac{3p_{0}}{m_{f}n_{0f}}-\frac{8p_{0f}g_{2f}m_{f}}{\hbar^{2}})^{3}}\Biggr],\end{equation}
and
$$\tilde{c}=g_{2b}n_{0b} -\frac{\hbar^{2}}{4m_{b}} +g_{2bf}n_{0b}\frac{\frac{n_{0f}g_{bf}}{m_{f}}}{V^{2} -\frac{3p_{0}}{m_{f}n_{0f}}-\frac{8p_{0f}g_{2f}m_{f}}{\hbar^{2}}}$$
\begin{equation} \label{BLS20 c tilde}
-\frac{\hbar^{2}}{4m_{f}}\frac{\frac{n_{0f}g_{bf}}{m_{f}}\frac{n_{0b}g_{bf}}{m_{f}}}{(V^{2} -\frac{3p_{0}}{m_{f}n_{0f}}-\frac{8p_{0f}g_{2f}m_{f}}{\hbar^{2}})^{2}}.\end{equation}
Coefficient $\tilde{a}$ is always positive since $V^{2}$ is positive for solution to exist.
However, condition $V^{2}>0$ gives a restriction on parameters.
For instance if we drop contribution of fermions $V^{2}=g_{b}n_{0b}/m_{b}$.
Hence, the interaction between bosons should be repulsive $g_{b}>0$.

Equation (\ref{BLS20 KdV}) can be reduced to single variable after introduction of new variable $\zeta=\tau-U\xi$.
Afterwards KdV equation can be integrated.
As the result of integration find nonlinear perturbation of boson concentration in the first order:
\begin{equation} \label{BLS20 solution for n 1 b} n_{1b}=\frac{3U\tilde{a}}{\tilde{b}} \frac{1}{\cosh^{2}\biggl(\frac{1}{2}\sqrt{\frac{U\tilde{a}}{\tilde{c}}}\zeta\biggr)}.\end{equation}
Since coefficient $\tilde{a}$ is positive, solution (\ref{BLS20 solution for n 1 b}) can exist if coefficient $\tilde{c}$ is positive.
Sign of coefficient $\tilde{b}$ defines the type of soliton:
bright soliton for $\tilde{b}>0$ or the dark soliton for $\tilde{b}<0$.

\begin{figure}
\includegraphics[width=8cm,angle=0]{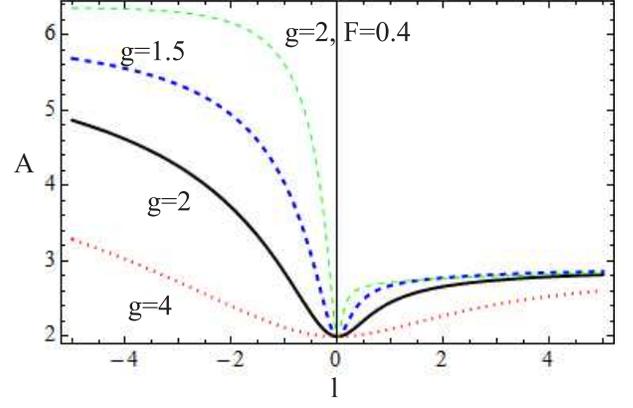}
\caption{\label{BLS20 Fig amplitude}
Dimensionless amplitude of soliton (\ref{BLS20 KdV}) $A=3Ua/bn_{0b}$
Three lower lines is plotted for $F=0.1$.
The upper line is made for $F=0.4$.}
\end{figure}

\begin{figure}
\includegraphics[width=8cm,angle=0]{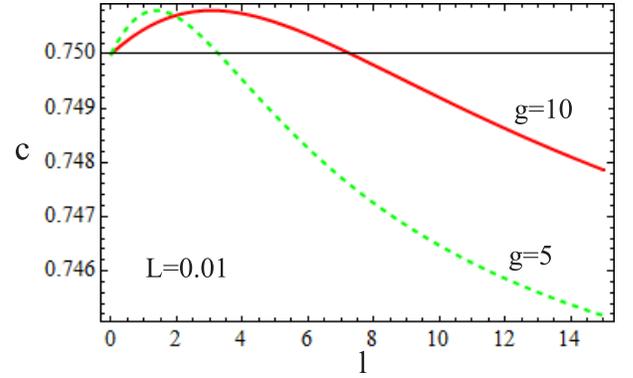}
\caption{\label{BLS20 Fig c l}
Dimensionless coefficient $c$ is demonstrated as function of dimensionless boson-fermion interaction $l$ at fixed $L$
for two values of boson-boson interaction constant $g=5$ and $g=10$ demonstrated in the figure.
Other parameters have the following values: $M_{0}=4$, $N_{0}=2$, $F=0.01$, $G=1$, $L=0.01$.}
\end{figure}

\begin{figure}
\includegraphics[width=8cm,angle=0]{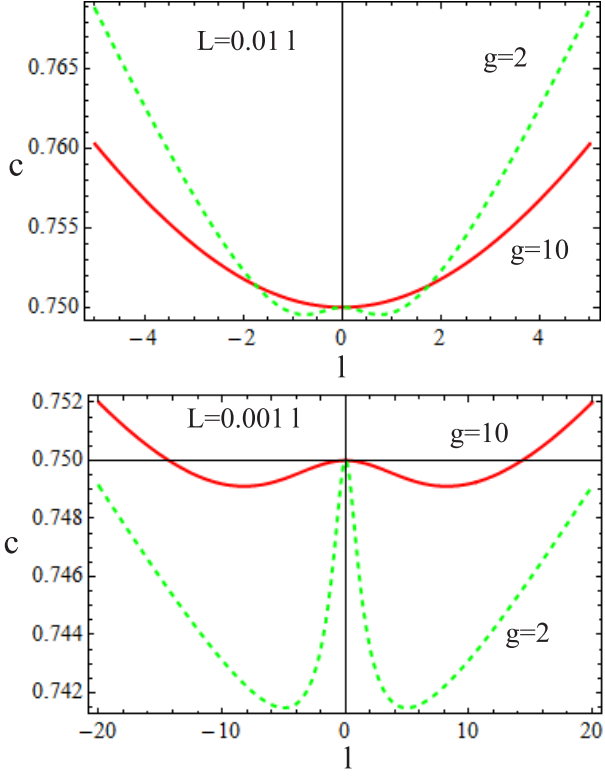}
\caption{\label{BLS20 Fig  c l L}
Dimensionless coefficient $c$ is presented as function of $l$ at simultaneous change of $L$ as $L=\alpha_{L}l$
for two values of $\alpha_{L}$: $\alpha_{L}=0.01$ and $\alpha_{L}=0.001$.
Each of them is given for two values of boson-boson interaction constant $g=2$ and $g=10$.
Other parameters have the following values: $M_{0}=4$, $N_{0}=2$, $F=0.01$, $G=1$.}
\end{figure}

\begin{figure}
\includegraphics[width=8cm,angle=0]{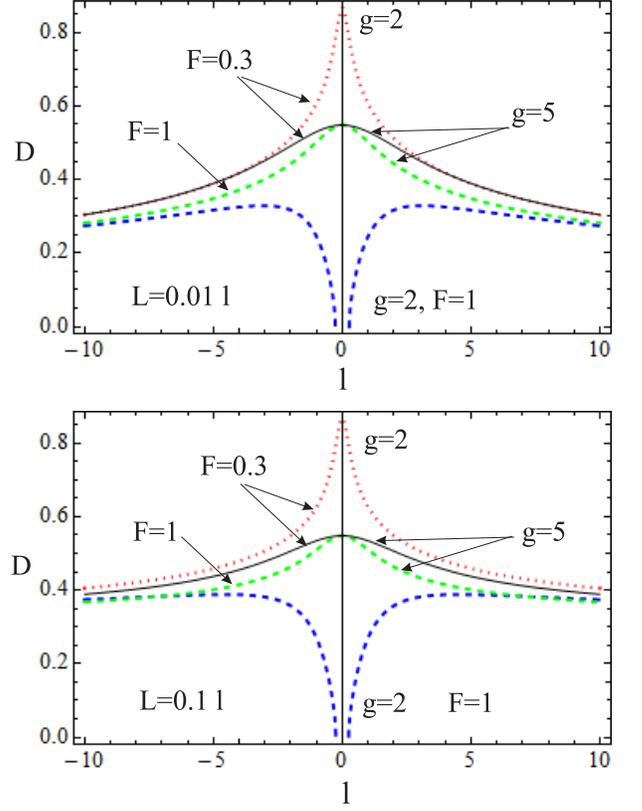}
\caption{\label{BLS20 Fig width 1}
Dimensionless width of soliton (\ref{BLS20 KdV}) $D=2\sqrt{c}/n_{0b}^{1/3}\sqrt{U a}$.
Other parameters have the following values: $M_{0}=4$, $N_{0}=2$, $G=1$.
Values of parameters $g$, $L$ and $F$ are given in the figure.}
\end{figure}

Purely for bosons the coefficient $\tilde{c}$ consists of two terms $\tilde{c}_{b}=g_{2b}n_{0b}-\hbar^{2}/4m_{b}^{2}$.
Hence, it is positive for nonzero interaction between bosons in third order by the interaction radius.
Moreover, it requires the repulsive boson-boson interaction $g_{2b}>0$.
This conclusion is in agreement with condition $V^{2}>0$ requiring $g_{b}>0$.
For bosons, coefficient $\tilde{a}$ simplifies to $\tilde{a}_{b}=2m_{b}V^{2}>0$.
Coefficient $\tilde{b}$ also appears in simple form in this limit:
$\tilde{b}_{b}=g_{b}+2m_{b}V^{2}/n_{0b}>0$.
The transition to bosons is made by limit $g_{bf}\rightarrow 0$.
It shows existence of a specific soliton solution with positive amplitude in system of bosons studied up to the TOIR.
Terms proportional to $g_{bf}$ give contribution of fermions
which is discussed numerically below.

This bright soliton solution purely for bosons is obtained in Ref. \cite{Andreev MPL B 12} and \cite{Andreev LP 19}.
Physical picture behind the bosonic bright soliton solution for repulsive bosons demonstrates
deep relation to the soliton solution experimentally obtained in Ref. \cite{Wang NJP 14}.

\section{Numerical analysis of the bright-like soliton}

To perform analysis of the soliton properties and area of its existence represent KdV equation (\ref{BLS20 KdV})
and velocity of perturbation (\ref{BLS20 solution for V}) in dimensionless form.

The dimensionless velocity is given via the mass of bosons and concentration of bosons
since we consider the soliton
which originally exists in the boson subsystem:
$$W^{2}\equiv\frac{m_{b}^{2}V^2}{\hbar^{2}n_{0b}^{2/3}}
=\frac{1}{2}\biggl\{g+\frac{3}{5}(6\pi^{2})^{\frac{2}{3}}\frac{N_{0}^{\frac{2}{3}}}{M_{0}^{2}}\biggl(1+\frac{8}{3}F\biggr)$$
\begin{equation} \label{BLS20 W-2} +\sqrt{\biggl(g-\frac{3}{5}(6\pi^{2})^{\frac{2}{3}}\frac{N_{0}^{\frac{2}{3}}}{M_{0}^{2}}\biggl(1+\frac{8}{3}F\biggr)\biggr)^{2} +4\frac{N_{0}}{M_{0}}l^{2}}\biggr\}, \end{equation}
where
$N_{0}=n_{0f}/n_{0b}$,
$M_{0}=m_{f}/m_{b}$,
$g\equiv m_{b}g_{b}n_{0b}^{1/3}/\hbar^{2}$,
$G\equiv m_{b}g_{2b}n_{0b}/\hbar^{2}$,
$l\equiv m_{b}g_{bf}n_{0b}^{1/3}/\hbar^{2}$,
$L\equiv m_{b}g_{2bf}n_{0b}/\hbar^{2}$,
and
$F\equiv m_{f}g_{2f}n_{0f}/\hbar^{2}$.
The explicit form of the equilibrium pressure for the degenerate fermions
with the full spin polarization is used in form of the Fermi pressure:
$p_{0f}=(6\pi^{2})^{2/3}\hbar^{2}n_{0f}^{5/3}/5m_{f}$.

Dimensionless KdV can be written as follows
\begin{equation} \label{BLS20 KdV dimless}
a\partial_{\tilde{\tau}}N+bN\partial_{\tilde{\xi}}N+c\partial_{\tilde{\xi}}^{3}N=0,\end{equation}
where $N=n_{1b}/n_{0b}$, $\tilde{\tau}=\sqrt[3]{n_{0b}}\tau$, $\tilde{\xi}=\sqrt[3]{n_{0b}}\xi$,
\begin{equation} \label{BLS20 hat a}
a=\frac{m_{b}\tilde{a}}{\hbar^{2}n_{0b}^{\frac{2}{3}}}=2W^{2}\biggl[1+\frac{N_{0}}{M_{0}}\frac{l^{2}}{D_{W}^{2}}\biggr], \end{equation}
$$b=\frac{m_{b}\tilde{b}}{\hbar^{2}n_{0b}^{\frac{1}{3}}}=2W^{2}+g
+\frac{N_{0}}{M_{0}}\frac{l^{2}}{D_{W}}
+\frac{N_{0}}{M_{0}^{2}}\frac{l^{3}}{D_{W}^{2}}$$
\begin{equation} \label{BLS20 hat b}
+2\frac{N_{0}}{M_{0}^{2}}\frac{l^{3}}{D_{W}^{3}}
\biggl[W^{2}+\frac{3}{5}(6\pi^{2})^{\frac{2}{3}}\frac{N_{0}^{\frac{2}{3}}}{M_{0}^{2}}\biggl(1+\frac{10}{3}F\biggr)\biggr],\end{equation}
and
\begin{equation} \label{BLS20 hat c}
c=\frac{m_{b}\tilde{c}}{\hbar^{2}}
=-\frac{1}{4}+G +\frac{N_{0}}{M_{0}}L\frac{l}{D_{W}}
-\frac{1}{4}\frac{N_{0}}{M_{0}^{3}} \frac{l^{2}}{D_{W}^{2}} ,\end{equation}
with
\begin{equation} \label{BLS20 def D W}D_{W}=W^{2}-\frac{3}{5}(6\pi^{2})^{\frac{2}{3}}\frac{N_{0}^{\frac{2}{3}}}{M_{0}^{2}}\biggl[1+\frac{8}{3}F\biggr] .\end{equation}

Change of sign of $g_{bf}$ does not change the velocity of perturbation $V$ (\ref{BLS20 solution for V}) and coefficient $\tilde{a}$ (\ref{BLS20 a tilde}).

Change of sign of $g_{bf}$ does not affect coefficient $\tilde{c}$ (\ref{BLS20 c tilde}) since it is also required change of sign of $g_{2bf}$.
While coefficient $\tilde{c}$ contains terms without $g_{bf}$ and $g_{2bf}$, term with square of $g_{bf}$, and term with product of $g_{bf}$ on $g_{2bf}$.

Change of sign of $g_{bf}$ modifies coefficient $\tilde{b}$ (\ref{BLS20 b tilde}).
If we consider parameters chosen in Fig. (\ref{BLS20 Fig form of n b 1}) find
that for $g=1$ change of sign of $g_{bf}$ modifies coefficient $\hat{b}$ in two times: $\hat{b}_{+}/\hat{b}_{-}\approx2$.
However, if parameter $g$ equals to $3$ or more change of sign of $g_{bf}$ gives few percent modification of $\hat{b}$.

Influence of the boson-boson interaction via $g$ on the form of soliton is demonstrated in Fig. (\ref{BLS20 Fig form of n b 1}).
Existence of the soliton requires relatively large boson-boson interaction to get $G>1/4$.
However, Fig. (\ref{BLS20 Fig form of n b 1}) shows that increase of boson-boson interaction
(change of $g$ from 1 to 10 at fixed $G=1$) leads to decrease of amplitude and width of the soliton.
Further increase of $g$ at fixed $G=1$ leads to decrease of width with no change of amplitude.
However, formal increase of $g$ up to $100$ together with increase of $G$ up to $G=10$ increases the width with no modification of amplitude.

The Fig. (\ref{BLS20 Fig form of n b 1}) is made for relatively small influence of the fermions.
Increase the contribution of fermions and find their contribution in properties of perturbations.

\textit{First focus on properties of the velocity.}
Dependence of the dimensionless velocity (\ref{BLS20 W-2}) on the boson-boson interaction $g$
for different boson-boson $l$ and fermion-fermion $F$ interactions is presented in
Figs. (\ref{BLS20 Fig W 1}), (\ref{BLS20 Fig W 2}), (\ref{BLS20 Fig W 3}).
Each figure is made for different rations of masses $M_{0}$ and concentrations $N_{0}$.

Velocity square $W^{2}$ (\ref{BLS20 W-2}) shows almost linear dependence on the boson-boson interaction $g$.
However, the presence of fermions change this dependence from linear to the superposition of the linear and the square root functions.

Main change of the dependence happens at small boson-boson interaction.
Figs. (\ref{BLS20 Fig W 1}), (\ref{BLS20 Fig W 2}), (\ref{BLS20 Fig W 3}) are made for relatively strong boson-fermion interaction $l\sim1$.
The fermion-fermion interaction is considered in an interval from average $F=0.1$ to strong $F=1$ values.
Mass (concentration) increase of each species decreases (increases) the velocity $V$ (\ref{BLS20 solution for V}).
This tendency conserves for dimensionless velocity (\ref{BLS20 W-2}),
where increase of the mass (concentration) of fermions relatively the boson mass (concentration) decreases (increases) the velocity $W$
(compare corresponding lines in Figs. (\ref{BLS20 Fig W 1}), (\ref{BLS20 Fig W 2}), (\ref{BLS20 Fig W 3})).
Relatively small influence of fermions is demonstrated in Fig. (\ref{BLS20 Fig W 1}),
where mass of fermions is relatively large $M_{0}=9$
while variation of $F$ is noticeable at small $g$.
The increase of $l$ increases the velocity $W$ as it is seen from analytical dependence (\ref{BLS20 W-2}).
The increase of fermion-fermion repulsion gives small increase of the velocity $W$ at small $g$ at fixed $l$.
Role of fermion-fermion interaction increases if the mass and concentration ratios are getting closer to $1$
as it is demonstrated at transition to Figs. (\ref{BLS20 Fig W 2}) and (\ref{BLS20 Fig W 3}).

\textit{On coefficients $b$ and $c$.}
Coefficient $c$ is  the symmetric function of $l$ for the fixed $\alpha_{L}$.
But coefficient $b$ shows nonsymmetric dependence on boson-fermion interaction $l$.
The third term in (\ref{BLS20 hat b}) is positive since $D_{W}>0$.
However, the fourth and last terms can be negative for attractive boson-fermion interaction.

Hence, the boson-fermion repulsion increases the amplitude of soliton.
The increase can be nonmonotonic
since $W^{2}(l^{2})$ are located in the denominator of the amplitude.

There is a competition between different terms defining the amplitude $A=3Ua/bn_{0b}$ for the attraction between bosons and fermions.
The fourth and last terms become negative in this regime
while the third term is positive.
Hence, the sign of the amplitude change depends on parameters of the system (see Fig. \ref{BLS20 Fig amplitude}).

Area of the soliton existence is restricted by the condition that
the width of soliton $D\sim\sqrt{c}/\sqrt{a}$ is real.
It means that coefficient $c$ should be positive, since coefficient $a>0$ is positive for all parameters.

There is simple dependence of the width of soliton on $G$.
It is linear via the second term in $c$.
Focus on $G=1$.

The third term in $c$ contains dependence on $L$.
This is positive term for positive product $lL$.
Let us to point out that $D_{W}$ is positive for all parameters.
Since parameters $l$ and $L$ are related introduce the following relation $L=\alpha_{L} l$,
where $\alpha_{L}<1$ is a parameter
which does not depend on $l$ or other parameters
and represents independent variation of interaction constant $L$.

Coefficient $\alpha_{L}$ is an independent parameter.
Hence, if $L$ is fixed at change of $l$.
It means that parameter $\alpha_{L}$ changes to compensate contribution of $l$ in $L$.

If we consider dependence of $c$ on $l$ at fixed $L$ the third term in $c$ (\ref{BLS20 hat c}) plays crucial role
(for instance at $m = 4$, $n = 2$, $g = 10$, $F = 0.01$, $G = 1$, $L = 0.01$).
The dependence numerically appears as almost parabolic dependence (see Fig. \ref{BLS20 Fig c l})
in spite more complex analytical dependence via $W^{2}(l)$.
This parabola has branches going below from the maximum located at positive value of $l$.

However, $l$ and $L$ are moments of the same potential of boson-fermion interaction.
So we use representation $L=\alpha_{L} l$ introduced above.
It changes dependence of $c$ on $l$.
In this case, parameter $c$ is the function of $l^{2}$.

Value $G=1$ is chosen, so the boson-fermion interaction shifts coefficient $c$ from value $c_{0}=0.75$.
For small positive $l$ at fixed $L=0.01$ the shift of $c$ is positive (see Fig. \ref{BLS20 Fig c l}).
There is value of $l=l_{0}(g)$,
where the shift becomes equal to zero.
Value of $l_{0}$ becomes larger at larger boson-boson interaction $g$.
At further increase of $l$ above $l_{0}$ the shift becomes negative $c<0.75$.
However, parameter $c$ shows small deviation from $G-1/4$
and has positive value.
Therefore, presence of fermions does not destroy the soliton solution.

Figs. \ref{BLS20 Fig c l} and \ref{BLS20 Fig c l L} show that
deviations of $c$ from value $G-0.25$ are small.
Therefore, small values of $G$ can be chosen down to $G_{min}=0.26$.

Consider behavior of $c$ at fixed $\alpha_{L}$.
Monotonic increase of $c$ as function of $l^{2}$ is found at relatively large $\alpha_{L}=0.01$
and relatively large $g=10$.
Small $g=2$ at large $\alpha_{L}=0.01$ and different $g$ at smaller $\alpha_{L}=0.001$ lead to decrease of $c$ at small $l$
which replaces by the increase of $c$ at larger $l$.
The area of decrease of $c$ from $c_{0}=0.75$ becomes wider and $c_{min}$ becomes smaller at smaller $\alpha_{L}$ and smaller $g$
as it is presented in Fig. \ref{BLS20 Fig  c l L}.
All of it is obtained for small fermion-fermion interaction $F=0.01$.
Area of larger $F$ is presented for the width of soliton $D$ on Fig. \ref{BLS20 Fig width 1}.

Relatively large fermion concentration and large repulsive fermion-fermion interaction can significantly decrease $D_{W}$.
So, the contribution of the last negative term in $c$ (\ref{BLS20 hat c}) can increase faster in compare with the third term in coefficient $c$.

Consider two parts of Fig. \ref{BLS20 Fig width 1}.
If $F=1$ there is no visible modification of $D(l)$ at different $\alpha_{L}$.
If the fermion-fermion repulsion is smaller $F=0.3$ there is increase of function $D(l)$ with increase of $\alpha_{L}$ at $g=2$.
However, the stronger boson-boson repulsion hides any contribution of $\alpha_{L}$.

Crucial role in the soliton existence plays the second constant of the boson-boson interaction $G>0.25$.
First interaction constants for the boson-boson and boson-fermion interactions defines properties of the solution.
The constants of the boson-fermion and fermion-fermion interaction existing in the third order on the interaction radius have small influence
if the boson-boson repulsive interaction is strong $g\geq1$.
Condition $G>0.25$ also corresponds to this criterium.

\section{Conclusion}

Boson-fermion mixtures have been studied in terms of hydrodynamic model.
Boson-boson, fermion-fermion and boson-fermion interactions have been considered up to the third order by the interaction radius.
A stress has been made on the models of fermions,
where the pressure tensor has been considered as an independent function.
Hence, no equation of state has been used for perturbations of pressure,
but additional hydrodynamic equation for the pressure evolution is derived from the microscopic quantum model.
Equation of state can be used for the equilibrium pressure.

Developed model has been used to study the bright soliton in repulsing Bose-Einstein condensate fraction.
It exists due to the repulsive boson-boson interaction giving positive interaction constant in the third order by the interaction radius.
Formation of soliton in fermion fraction has been found.
It has been obtained that type of soliton of fermion concentration depends
on the sign of boson-fermion interaction constant in the first order by the interaction radius.
Hence, the boson-fermion repulsion (attraction) leads to bright (dark) soliton in fermion fraction.
Influence of the fermions on the properties of soliton in boson fraction is analyzed.

The obtained model contains the first order on the interaction radius including the boson-boson interaction corresponding to the Gross-Pitaevskii equation
and boson-fermion interaction (existing in well-known and sited above works on boson-fermion mixtures)
which are three-dimensional zeroth moments of the interaction potential.
However, the consideration of the interaction terms in the third order by the interaction radius introduces three additional interaction constants
which are the second moments of the interaction potential for boson-boson, boson-fermion, and fermion-fermion interactions.

It is possible to make an estimation of new constants via well-known constants (for boson-boson and boson-fermion interactions) as it is presented in \cite{Andreev PRA08}.
However, all constants are independent and introduce additional information about interaction potential.
Hence, the experimental study of properties of the found here solitons allows to study the interaction potential in more details.

Moreover, the found solitons presents interest by themselves
since they are examples of new nonlinear phenomena in ultracold mixtures.

\section{Acknowledgements}
The work of P.A. and M.T. is supported by the Russian Foundation for Basic Research (grant no. 20-02-00476).

\end{document}